\documentclass[letterpaper, 10 pt, journal, twoside]{IEEEtran}  

\IEEEoverridecommandlockouts                              
\usepackage{graphicx} 
\usepackage{amsmath} 
\usepackage{amssymb}  
\usepackage[usenames, dvipsnames]{color}

\usepackage{lipsum}
\usepackage{color}
\usepackage{cite}

\usepackage{diagbox} 
\usepackage{tabu}

\usepackage[ruled,linesnumbered]{algorithm2e}
\usepackage{algpseudocode}
\usepackage{epsfig}
\usepackage{booktabs}  
\usepackage{multirow}  

\newcommand{\tabincell}[2]{\begin{tabular}{@{}#1@{}}#2\end{tabular}}

\newcommand{\revision}[1]{\textcolor{black}{#1}} 

\title{Deformation-Aware Robotic 3D Ultrasound
}

\author{Zhongliang Jiang$^{1}$, Yue Zhou$^{1}$, Yuan Bi$^{1}$, Mingchuan Zhou$^{2}$, Thomas Wendler$^{1}$, and Nassir Navab$^{1,3}$ 

\thanks{
\textit{(Zhongliang Jiang and Yue Zhou contributed equally to this work). (Corresponding author: Mingchuan Zhou.)}}

\thanks{$^{1}$Zhongliang Jiang, Yue Zhou, Yuan Bi, Thomas Wendler and Nassir Navab are with the Chair for Computer Aided Medical Procedures and Augmented Reality, Technical University of Munich, Germany. {\tt\footnotesize{(zl.jiang@tum.de)}}
        }%
        
\thanks{$^{2}$Mingchuan Zhou is with the College of Biosystems of Engineering and Food Science, Zhejiang University, China.}

\thanks{$^{3}$Nassir Navab is also with the Laboratory for Computational Sensing and Robotics, Johns Hopkins University, Baltimore, USA.}

}

\begin{document}

\maketitle

\begin{abstract}
Tissue deformation in ultrasound (US) imaging leads to geometrical errors when measuring tissues due to the pressure exerted by probes. 
Such deformation has an even larger effect on 3D US volumes as the correct compounding is limited by the inconsistent location and geometry. 
This work proposes a patient-specified stiffness-based method to correct the tissue deformations in robotic 3D US acquisitions. 
To obtain the patient-specified model, robotic palpation is performed at sampling positions on the tissue. The contact force, US images and the probe poses of the palpation procedure are recorded. The contact force and the probe poses are used to estimate the nonlinear tissue stiffness. The images are fed to an optical flow algorithm to compute the pixel displacement. Then the pixel-wise tissue deformation under different forces is characterized by a coupled quadratic regression. To correct the deformation at unseen positions on the trajectory for building 3D volumes, an interpolation is performed based on the stiffness values computed at the sampling positions. With the stiffness and recorded force, the tissue displacement could be corrected. The method was validated on two blood vessel phantoms with different stiffness. The results demonstrate that the method can effectively correct the force-induced deformation and finally generate 3D tissue geometries.
\end{abstract}

\markboth{IEEE Robotics and Automation Letters. Preprint Version. Accepted July, 2021}
{Jiang \MakeLowercase{\textit{et al.}}: Deformation-Aware Robotic 3D Ultrasound}

\begin{IEEEkeywords}
Medical Robots and Systems; Robotic Ultrasound; US deformation correction; 3D ultrasound
\end{IEEEkeywords}



\bstctlcite{IEEEexample:BSTcontrol}
\section{Introduction}
\begin{figure}[ht!]
\centering
\includegraphics[width=0.45\textwidth]{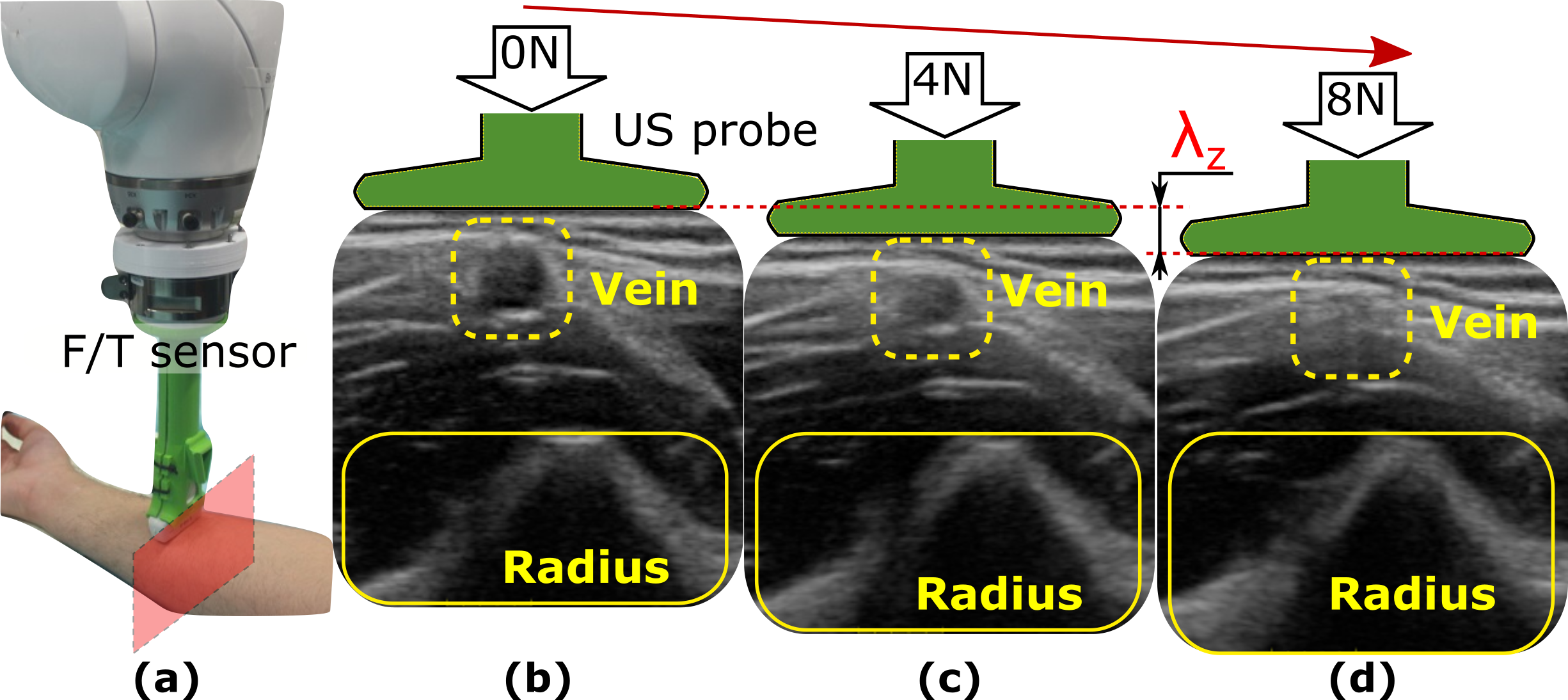}
\caption{Cephalic vein imaged under three different forces. (a) robotic arm with a force/torque (F/T) sensor, (b), (c) and (d) are the resulting B-mode images acquired when the contact forces are $0~N$, $4~N$ and $8~N$, respectively. The echo of the radius bone is visible below. $\lambda_z$ represents probe tip displacement in the applied force direction between (b) and (d), here about $6~mm$. 
}
\label{Fig_defromation_background}
\end{figure}

\IEEEPARstart{U}{ltrasound} (US) is a broadly utilized diagnostic imaging modality for examinations of internal organs. It is also commonly used to obtain the location and geometric information of disease intraoperatively as US imaging is \revision{highly available}, non-invasive and radiation-free. However, to obtain optimal acoustic coupling of a US transducer and thus achieve good visibility of target anatomies, a certain pressure is required to be applied to the imaged anatomy. Due to the exerted pressure, the shape distortion of visualized tissue structures is inevitable, particularly for soft tissues such as superficial blood vessels (Fig.~\ref{Fig_defromation_background}). The shape of the cephalic vein continues to compress when the contact force increases. The vein loses its complete lumen when the force increases to $8~N$. \revision{As a result, the distortion can severely obfuscate the geometrical measurements of subsurface targets, e.g., measuring blood vessel diameter for diagnosing vascular stenosis. }

It is challenging to maintain a constant force even for an experienced sonographer during the traditional free-hand US acquisition. The varied pressure results in different deformations across the individual 2D US slices, which will further hinder the achievement of accurate 3D geometry. Although 3D US imaging can directly display 3D anatomical structures reducing the requirements for sonographers and improving the accuracy of diagnosis~\cite{pfister2016standardized}, the deformation impairs the higher acceptance of such approaches in clinical practice~\cite{virga2018use}.

\par
To address the pressure-induced deformation, Treece~\emph{et al.} combined non-rigid image-based registration and position sensing for free-hand 3D musculoskeletal US examinations~\cite{treece2002correction}. This method mainly focused on axial deformation. Burcher~\emph{et al.} built a finite element method (FEM) model to predict the deformation in both axial and lateral directions~\cite{burcher2001deformation}. The performance of this method relies on the prior knowledge of tested tissues, which makes it unsuitable for \revision{real-world} application. Additionally, Flach~\emph{et al.} employed a generic homogeneous representation of tissues to estimate the deformation using FEM~\cite{flack2016model}. However, tissues are inhomogeneous in reality and their properties vary from one patient to another. To address this drawback, Dahmani~\emph{et al.} applied a linear elastic biomechanical model to estimate the personalized mechanical parameters of the tissues along the deformation field~\cite{dahmani2017model}. In addition, Sun~\emph{et al.} proposed a method to achieve a zero-pressure image using an empirical regressive model of US image deformation with respect to the applied force~\cite{sun2010trajectory}. Nevertheless, the method was developed to correct the deformation in 2D \revision{slice}. A good regressive model is computed based on a set of forces and corresponding pixel displacements. Thus, this method cannot be directly extended to obtain compression-free 3D volumes because it is impractical to acquire paired forces and image deformations in a dense sampling.

\par
Recently, since robotic manipulators are much more accurate and stable than human operators, robotic US systems (RUSS) have become a potential solution toward automatic US scans~\cite{jiang2020automatic,gilbertson2015force,jiang2021autonomous,huang2018robotic}. To accurately control the probe orientation, Jiang~\emph{et al.} developed a method to align the probe centerline with the normal direction of a tested object using force data~\cite{jiang2020automatic_tie}. Besides, Jiang~\emph{et al.} proposed a motion-aware 3D RUSS to combine the flexibility of the free-hand US and the accuracy and stability of robot~\cite{jiang2021motion}. 
In addition, unlike the free-hand manner, the probe pressure-induced indentation to the tissues is accurately controlled to be homogeneous along the entire US volume for RUSS. This is achieved by using hybrid force/position controllers to maintain a certain contact force during the scanning~\cite{gilbertson2015force}. However, homogeneous deformations still exist in all slices. To further obtain zero-pressure 3D volumes, Virga~\emph{et al.} applied a 4th-order polynomial function to regress the force-dependent deformations and further propagate the deformation field at sparse sampling points to the whole sweep direction~\cite{virga2018use}. Nevertheless, five pixels are required to be manually selected on the first frame. In addition, the method takes around $186~s$ to compute the deformation field at one location and 15 points need to be selected along a sweep path of $70~mm$. These drawbacks hinder the use of this method in real clinical practice. 

\par
In this work, a stiffness-based deformation correction method, incorporating image pixel displacements, contact forces and nonlinear tissue stiffness, is proposed to recover a zero-compression 3D tissue geometry from the deformed data recorded during robotic scans. To obtain patient-specific regression models, robotic palpation was performed at sampling positions. Since the tissue compression leads to the increase of tissue stiffness, the nonlinear tissue stiffness was modeled as a 2nd-order polynomial function to the probe displacement in the applied force direction. An optical flow algorithm was employed to compute the two-dimensional pixel displacement in US images acquired under different contact forces. With such displacement, a coupled quadratic regression model with respect to the pixel position, the contact force and the tissue stiffness was computed as an optimization issue for the sampling position. Since the tissue stiffness is the key factor affecting the deformation, the optimized regression model for the initial sampling position can be quickly propagated to other positions on the trajectory by substituting the estimated local stiffness. This speeds up the process to obtain compression-free 3D volumes. The method was validated on two vascular phantoms with significantly different stiffness.



  



\section{System Overview}

\subsection{Hardware Overview}
\begin{figure}[ht!]
\centering
\includegraphics[width=0.42\textwidth]{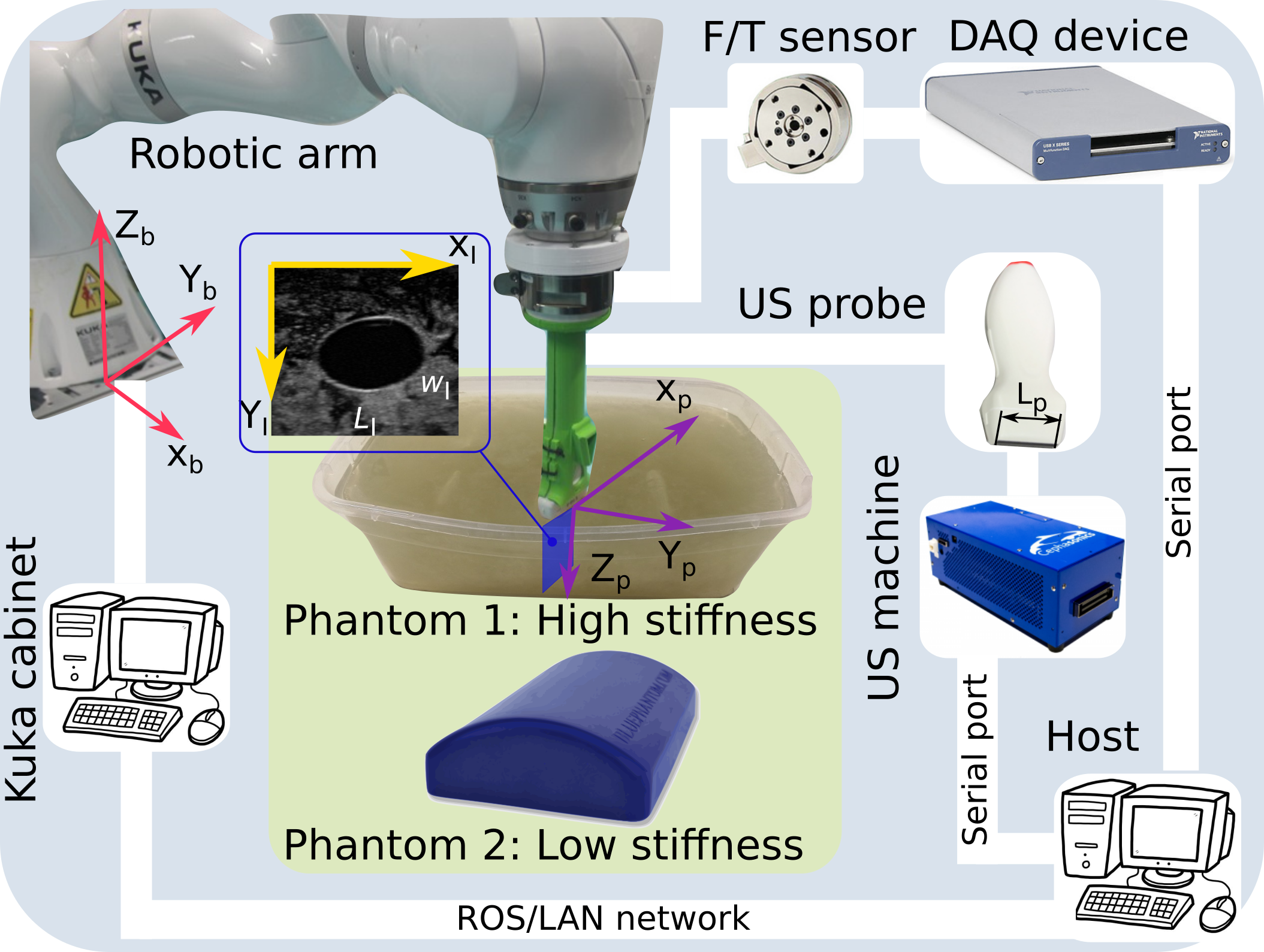}
\caption{System overview.}
\label{Fig_system_overview}
\end{figure}

\par
The system is comprised of three main components: a robotic manipulator (KUKA LBR iiwa 7 R800, KUKA Roboter GmbH, Germany), a linear US probe (CPLA12875, Cephasonics, USA) and a Gamma force/torque (F/T) sensor (ATI Industrial Automation, USA). The F/T sensor is attached to the robotic flange. The US probe is mounted on the other side of the F/T sensor using a custom-designed holder. The robotic system is controlled using a Robot Operating System (ROS) interface~\cite{hennersperger2016towards}. The control commands are exchanged at $100~Hz$. The whole system is depicted in Fig.~\ref{Fig_system_overview}.


\par
The contact force between the probe and the object is recorded using a data acquisition device (FTD-DAQ-USB6361, National Instruments, USA) and further published to ROS with accurate timestamps. Besides, the US images ($50~fps$) are accessed via a USB interface and visualized using a software platform (ImFusion GmbH, Munich, Germany). The detailed acquisition parameters are set as follows: image depth: $40~mm$, frequency: $7.6MHz$, brightness: $66~dB$. To synchronize the US images, forces, and probe poses, the US images are also published to ROS with timestamps.

\par
To validate the performance of the proposed method to recover zero-compression images from deformed images on different tissues, two blood vessel phantoms with different stiffness are employed (Fig.~\ref{Fig_system_overview}). The one with higher stiffness is custom-made by gelatin powder ($175~g/L$) and paper pulp ($3$-$5~g/L$). The paper pulp is used to mimic the unstructured artifacts of human tissues in US images. In addition, to increase the durability of the gel phantom (over one month), a liquid disinfectant is mixed with water ($1:9$). The second phantom (blue phantom, CAE, FL, USA) is softer than the custom-designed one. The phantoms with significantly different stiffness are used to validate whether the proposed method is able to quickly extrapolate its optimized correction model to different tissues without any prior knowledge. 


\subsection{System Calibration}

\par
To generate a 3D US volume from the tracked 2D B-scans, the spatial calibration should be employed to project the pixels to voxels in 3D space. There are three coordinate frames involved: robotic base frame $\{b\}$, US probe tip frame $\{p\}$, and B-mode imaging frame $\{I\}$ (see Fig.~\ref{Fig_system_overview}). Since $Z$ direction of frame $\{I\}$ is set in the perpendicular direction of the 2D image plane, a homogeneous representation of pixel position in frame $\{I\}$ can be written as $^{I}\textbf{P} = [x, y, 0, 1]^{T}$. Accordingly, 2D pixel positions $^{I}\textbf{P}$ are projected into 3D space using Eq.~(\ref{eq_trasformation}).

\begin{equation}\label{eq_trasformation}
^{b}\textbf{P} = ^{b}_{p}\textbf{T}~^{p}_{I}\textbf{T}~^{I}\textbf{P}
\end{equation}
where $^{j}_{i}\textbf{T}$ is the transformation matrix used to transfer the position from frame $\{i\}$ to frame $\{j\}$.

\par 
The transformation matrix $^{b}_{p}\textbf{T}$ is computed based on the kinematic model and the 3D geometry parameters of the 3D printed probe holder. In addition, considering that the origin of frame $\{p\}$ is set to the central point of the probe flat tip surface, and that the origin of frame $\{I\}$ is on the top left side, the matrix $^{p}_{I}\textbf{T}$ is represented as follows:

\begin{equation}\label{eq_I_to_probe}
^{p}_{I}\textbf{T} = \begin{bmatrix}
\frac{L_p}{L_I} & 0 & 0 & -\frac{L_p}{2}\\
0 & 0 & -1 & 0\\
0 & \frac{D_I}{W_I} & 0 & \varepsilon\\
0 & 0 & 0 & 1\\
\end{bmatrix}
\end{equation}
where $L_I$ and $W_I$ are the length and width (in pixels) of the 2D images \revision{(see Fig.~\ref{Fig_system_overview})}. $D_I$ is the physical depth of the US scan, namely $40~mm$. $L_p$ is the physical length of the deployed US elements, here $L_p = 37.5~mm$. $\varepsilon$ is used to represent the small distance from probe frame origin to image frame origin determined by the US elements configuration.   

\subsection{Control Architecture}
To ensure the image quality and human safety, a compliant controller is essential for robotic US acquisition~\cite{hennersperger2016towards}. The used Cartesian compliant control law is defined in Eq.~(\ref{eq_impedance_law})
\begin{equation}\label{eq_impedance_law}
\tau = J^{T}[K_m(x_d - x_c) + F_d] + D(d_m) + f_{dyn}(q,\Dot{q}, \Ddot{q})]
\end{equation}
where $\tau$ is the target torque applied to the joint drivers, $J^{T}$ is the transposed Jacobian matrix, $x_d$ and $F_d$ are the desired position and the target force, $x_c$ is current position, $K_m$ represents the Cartesian stiffness, $D(d_m)$ is the damping term, and $f_{dyn}(q,\Dot{q}, \Ddot{q})$ is the dynamic model of the robotic arm. 

\par
To make the force stable in the direction of the probe centerline, the robot is controlled via a 1-DOF compliant controller and a 5-DOF position controller. This behavior is achieved by assigning high stiffness values to the DOF that will be controlled under position mode. Regarding the compliant direction, the stiffness is set to a value in $[125, 500]~N/m$ for different human tissues~\cite{hennersperger2016towards}


\section{Pressure-Induced Deformation Correction}
\par
\begin{figure}[ht!]
\centering
\includegraphics[width=0.48\textwidth]{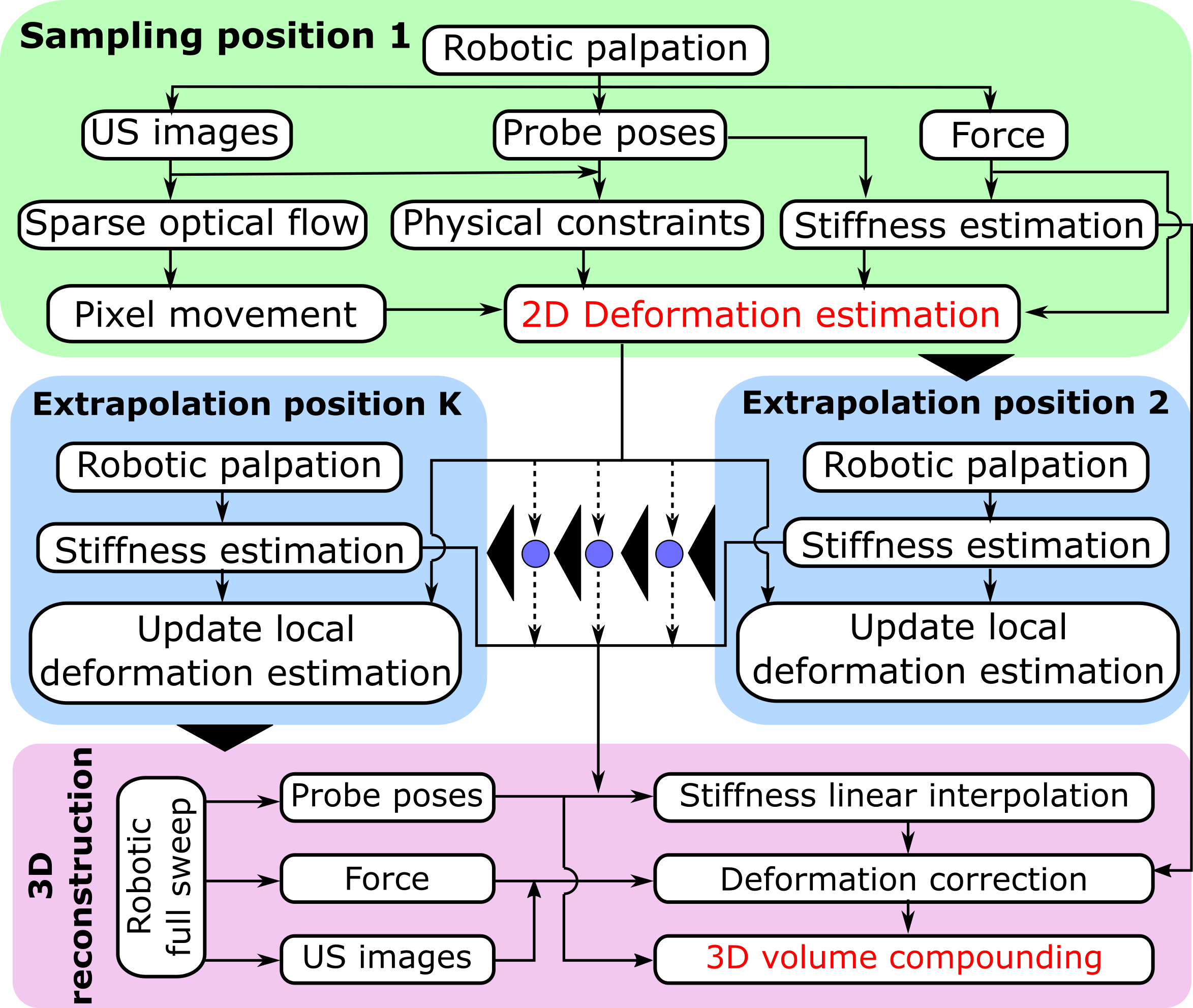}
\caption{Pipeline to correct the force-induced deformation and reconstruct 3D US volumes. At least two palpation procedures should be performed at different positions to obtain the zero-compression volume. 
}
\label{Fig_pipeline_correction}
\end{figure}

The overall pipeline proposed to compute compression-free 3D volumes based on multiple sensor information and a robotic platform has been shown in Fig.~\ref{Fig_pipeline_correction}. To overcome the limitation of the state-of-the-art method~\cite{virga2018use}, the proposed approach is developed based on the tissue stiffness. The use of stiffness allows to quickly propagate an estimated 2D deformation regression from one position to another position, or even extrapolate the optimized regression model to other tissues with totally different stiffness profiles. 

\par
Being beneficial from using a robotic platform, palpation can be automatically performed, and paired forces, probe poses and US images are recorded. Afterward, the local stiffness with respect to the contact force and tissue position is estimated. The unknown parameters of the regression model are optimized based on the pixel movements, extracted using optical flow technique, in the axial and lateral direction, respectively. Considering the tissue properties are not homogeneous, robotic palpation is performed multiple times at different positions. Instead of running multiple optimization procedures, the proposed method can update the deformation regression by substituting the local stiffness profiles in the optimized model. Finally, to obtain 3D volumes, a robotic sweep is performed along the planned path. The deformations of the US images acquired at unsampled positions are corrected by the approximated local stiffness computed based on the sampled positions with linear interpolation. Based on the corrected 2D images and probe poses, the 3D volume can be compounded. The detailed steps are described in the following subsections.

\subsection{2D Deformation Estimation}\label{Subsection_pixel_displacement}
\par
To accurately visualize the tissue geometry and enable better diagnosis, a 2D deformation field is computed using two regression functions with respect to tissue stiffness and coupling pixel displacements.
The tissues with high stiffness can resist deformations, while the tissues with low stiffness will suffer larger deformations during scans. However, most model-based deformation correction methods directly compute the pixel displacements with respect to the contact force ($F_c$) rather than to the stiffness ($k$)~\cite{treece2002correction, sun2010trajectory, virga2018use}. Since $F_c$ is applied externally, it cannot reflect tissue's deformation-resistant capacity. 
\revision{Therefore, the computed deformation field cannot be propagated on other positions or other tissues, which limits the force-based approach for building zero-compression 3D volumes.
}
\revision{
To propagate deformation fields for 3D volume, an additional 3D inpainting technique is required as in~\cite{virga2018use}. Yet, this technique takes too much time to compute the 2D deformation field. It reported that $15$ points are needed to build a good 3D volume of the sweep lengthen $70~mm$, and the computation for each point takes $186~s$ on average. Thus, here a stiffness-based displacement model is built. Since stiffness reflects tissue properties, the proposed approach allows to quickly adapt optimized regressions from sampled positions to unseen positions and even to other tissues. }

\subsubsection{Dynamic Tissue Stiffness Estimation} \label{stiffness}
\par
\begin{figure}[ht!]
\centering
\includegraphics[width=0.37\textwidth]{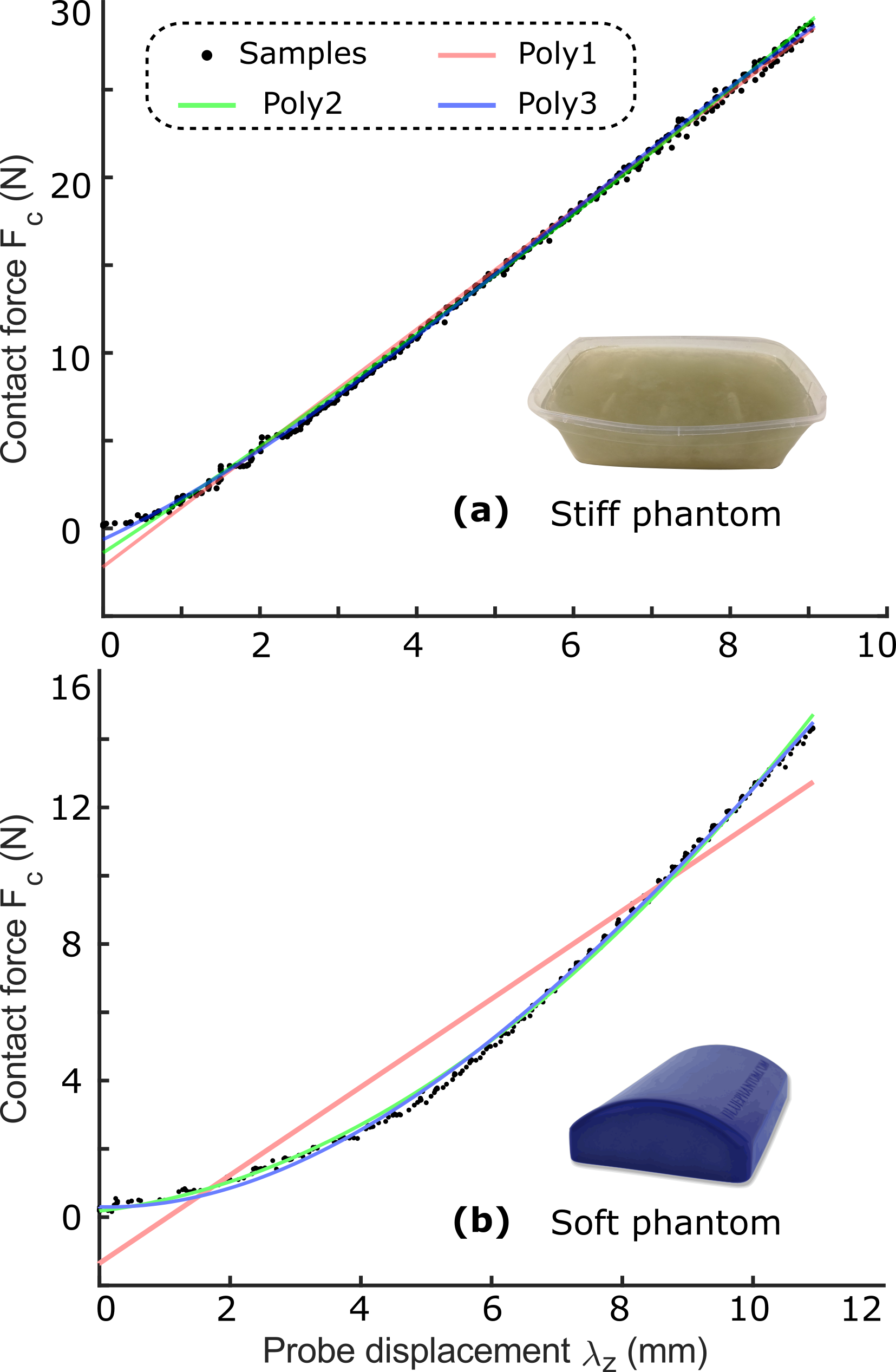}
\caption{Regression model for tissue stiffness. Poly1, Poly2 and Poly3 represent first, second and third order polynomial. The computed stiffness (Mean$\pm$SD) for the two phantoms are $3237\pm56~N/m$ and $1489\pm617~N/m$, respectively.}
\label{Fig_tissue_stiffness}
\end{figure}

Considering a real scenario, tissue stiffness also varies depending on the compression situation. The stiffness becomes large when large deformations occurred. As an example, the recorded force and probe tip displacement $\lambda_z$ (Fig.~\ref{Fig_defromation_background}) on the custom-designed stiff and the commercial soft phantom are shown in Fig.~\ref{Fig_tissue_stiffness}. During the robotic palpation procedure, the contact force is slowly increased $(0,30~N)$ for the stiff phantom and $(0, 16~N)$ for the soft one. It can be seen from Fig.~\ref{Fig_tissue_stiffness} that the samples acquired on the stiff phantom are distributed around a linear function while the samples acquired on the soft one are distributed around a nonlinear function. To qualitatively analyze the performance of the stiffness regression models, R-Squared ($R^2$) is measured. A quadratic function is already able to well explain the samples obtained on both stiff and soft tissues ($R^2 = 0.998$) while $R^2$ is only improved by $0.001$ using a cubic function. Therefore, a quadratic function is employed to capture the flexibility of elastic tissues as follows:

\begin{equation}\label{eq_displacement_force}
F_c = c_1\lambda_z^2 + c_2\lambda_z + c_3
\end{equation}
where $c_1$, $c_2$ and $c_3$ are the constant coefficients of the quadratic stiffness regression model. 

The dynamic stiffness $k_d$ according to current deformation is computed as the derivative of Eq.~(\ref{eq_displacement_force}).

\begin{equation}\label{eq_stiffness}
k_d = 2c_1\lambda_z + c_2
\end{equation}

\subsubsection{Pixel-wise Displacement Regression}
\par 
To characterize tissue deformation, pixel tracking is performed between US images acquired under different pressures using optical flow. Considering the accuracy and the cost of time, Lucas-Kanade approach~\cite{lucas1981iterative} is employed to estimate the pixel displacements in both lateral and axial directions $\textbf{D}_p^{op} = \left[{D}_{p}^{x}, {D}_{p}^{y} \right]^T$. 
Considering the pixel displacement $d_p$ is a coupled result over different factors, a high coupled polynomial regression over pixel position $(x,y)$, contact force $F_c$ and the estimated tissue stiffness $k_d$ are built. Since the deformation can be calculated using $F_c$ and $k_d$, $h=\frac{F_c}{k_d}$ is defined to estimate $d_p$ as follows:

\begin{equation}\label{eq_pixel_regression_constant_stiffness}
\textbf{d}_p(x,y,\frac{F_c}{k_d}) = \begin{bmatrix} \textbf{K}^x \\ \textbf{K}^y \end{bmatrix}~\textbf{M}_v(x,y,\frac{F_c}{k_d})
\end{equation}
where $\textbf{K}^x\in R^{1\times10}$ and $\textbf{K}^y\in R^{1\times10}$ are unknown parameters of the pixel regressions in lateral and axial direction, respectively.
$\textbf{M}_v(x,y,h)=[x^2,~y^2,~h^2,~xy,~xh,~yh,~x,~y,~h,~1]^T$ includes all variable combinations (2nd and 1st order) representing the coupling effect as~\cite{sun2010trajectory}. 


\par
However, the stiffness of soft tissues varies a lot with respect to the applied force as Fig.~\ref{Fig_tissue_stiffness}~(b). Eq.~(\ref{eq_pixel_regression_constant_stiffness}) is only able to depict the deformation of stiff tissues. To make the proposed method also work on soft tissues, the idea of calculus is employed to formalize the final regression model for each pixel as follows:

\begin{equation}\label{eq_pixel_regression_calculus}
\textbf{D}_p(F_c) = \int_{0}^{F_c}\textbf{d}_p(x,y,\frac{F}{k_d})dF
\end{equation}
where $\textbf{D}_p(F_c)$ is the cumulative result of $\textbf{d}_p$ computed for small force intervals, $k_d$ is seen as a constant in each interval.

\par
Besides the dynamic stiffness, this work also employs the probe displacement $\lambda_z$ (see Fig.~\ref{Fig_defromation_background}) as boundary constraints to better recover the geometry of the scanned tissue. With a larger $\lambda_z$, US imaging will visualize deeper tissues. This means the bottom pixels of the images acquired using small force will be moved toward the top when $F_c$ is increased. Due to the contact between the probe and object, the movement of the superficial layer is mainly in the axial direction. In addition, based on the optical flow results, the displacement of the pixels located in the low part also mainly happens in the axial direction, especially when a large force is applied. Thus, the physical constraints can be written as follows:

\begin{equation}\label{eq_physical_constraints}
\begin{split}
\textbf{D}_p^b(x,0, F_c) &= [0, 0]^T \\
\textbf{D}_p^b(x,L_I, F_c) &= [0, \frac{\lambda_z}{L_T}L_I]^T \\
\end{split}
~~~~~1\leq x \leq W_I
\end{equation}
where $L_I$ and $W_I$ represent the pixel size (length and width) of the US images, and $L_T$ is the thickness of the flexible layer. 

Then, the unknown parameters $\textbf{K}^x\in R^{1\times10}$ and $\textbf{K}^y\in R^{1\times10}$ can be optimized using Eq.~(\ref{eq_loss})

\begin{equation} \label{eq_loss}
\min_{\textbf{K}^x,~\textbf{K}^y}\frac{1}{MN}\sum_{i=1}^{M}\sum_{j=1}^{N}{||\textbf{D}_p(x^j,y^j,\frac{F_c^i}{k_d^i}) - \textbf{D}_p^m(x^j,y^j,F_c^i)||^2}
\end{equation}
where $\textbf{D}_p^m(x,y,F_c)$ is the measured pixel displacement. It consists of $\textbf{D}_p^b$ and $\textbf{D}_p^{op}$. $M$ is the number of paired force values and US images recorded during robotic palpation, $N$ is the number of the characterized pixel positions involved in optical flow and the boundary constraints. To make the regression can be extrapolated to other positions with different stiffness, $F_c$ and the pixel displacement $d_p^x$ and $d_p^y$ have been normalized to $[0,1]$. Besides, Eq.~(\ref{eq_loss}) is optimized using ADAM optimizer~\cite{kingma2014adam} with a updating step of $0.01$.

\subsection{3D Compression-Free Reconstruction}
\par
The pixel displacement regression model at an initial sampling position on the scanned tissue can be computed as described in Section~\ref{Subsection_pixel_displacement}. The US deformations at this position are computed by substituting the measured $F_c$ and $\lambda_z$ in the regression. With the computed pixel-wise displacements under different forces, the deformation can be compensated. To obtain the zero-compression volume, the correction procedures are carried out for the whole trajectory. To characterize the local feature, $N_k$ equidistant sampling positions are selected on the trajectory. Subsequently, the robotic palpation is performed at all sampling points and the corresponding stiffness can be estimated. 
Then, the deformation regression for a different position can be updated by substituting new stiffness.

\begin{equation}\label{eq_pixel_regression_new_position}
\textbf{D}_p(F_c) = \int_{0}^{F_c}\textbf{d}_p(x,y,\frac{F}{k_d^i})dF
\end{equation}
where $k_d^i$ is the nonlinear tissue stiffness with respect to the probe displacement $\lambda_z$ at $i$-th sampling position. 

\par
Since dense sampling is impractical for real scenarios, a robotic sweep over the anatomy is performed. Based on the probe poses, the local stiffness for unsampled positions on the trajectory is computed as follows: 

\begin{equation} \label{eq_stiffness_interpolation}
k_d^{'} = \sum_{i=0}^{N_k}\omega_k^i k_d^i
\end{equation}
where $\omega_k^i$ is weight for the stiffness at $N_k$ sampled positions, $\sum_{i=0}^{N_k}\omega_k^i  = 1$.
$\omega_k^i$ is determined using the distance $L_{d_i}$ from an unsampled position to all sampled positions as Eq.~(\ref{eq_stiffness_weight}).

\begin{equation} \label{eq_stiffness_weight}
\mathbf{\omega}_k=\left\{
\begin{aligned}
&\left[ \frac{\sum_{i=1}^{N_k}L_{d_i}}{L_{d_1}}, \dots, \frac{\sum_{i=1}^{N_k}L_{d_i}}{L_{dN_k}},  \right]_{Nor}  &\forall L_{d_j} \neq 0  \\
&[0,\dots, \underbrace{1}_{j-th}, \dots, 0 ]~~~~~~&\exists L_{d_j} = 0
\end{aligned}
\right.
\end{equation}

\par
Based on the measured force and the estimated stiffness in Eq.~(\ref{eq_stiffness_interpolation}), the deformations at unsampled positions can also be compensated using the optimized pixel regression for the initial position. The estimated compression-free volume for the target tissues is compounded using the paired corrected US images and the corresponding probe poses as~\cite{jiang2021motion}. 

\section{Results}
\subsection{2D Compensation Results}
\subsubsection{Validation of the Deformation Correction Method on the Sampling Position}
To validate the performance of the proposed method for generating zero-compression US images from deformed ones, robotic palpation was performed three times at the same position on the stiff phantom. The two sets of data are used to optimize the unknown regression parameters $\textbf{K}^x$ and $\textbf{K}^y$ in Eq.~(\ref{eq_pixel_regression_constant_stiffness}). The results are shown in TABLE~\ref{Table_optimized_parameters} (taking one arbitrary sampling point as an example).

\begin{table}[!ht]
\centering
\caption{Optimized Regression Parameters}
\label{Table_optimized_parameters}
\begin{tabular}{ccc}
\noalign{\hrule height 1.2pt}
Parameters  & Value & Loss\\ 
\noalign{\hrule height 1.0 pt}
$\textbf{K}^x$   & \tabincell{c}{$[-0.27, -0.01, 0.15, -0.25, -0.24,$\\ $-0.27, 0.29, 0.03, 0.16, -2.01]e^{-1}$} &  \multirow{2.5}{*}{$2.5e^{-5}$}\\

$\textbf{K}^y$  & \tabincell{c}{$[-1.43, -1.89, -4.62, -0.87, -0.81,$\\  $3.69, 0.80, -1.91, 2.29, -2.23]e^{-1}$} &  \\
\noalign{\hrule height 1.2 pt}
\end{tabular}
\end{table}



\par
\begin{figure}[ht!]
\centering
\includegraphics[width=0.4\textwidth]{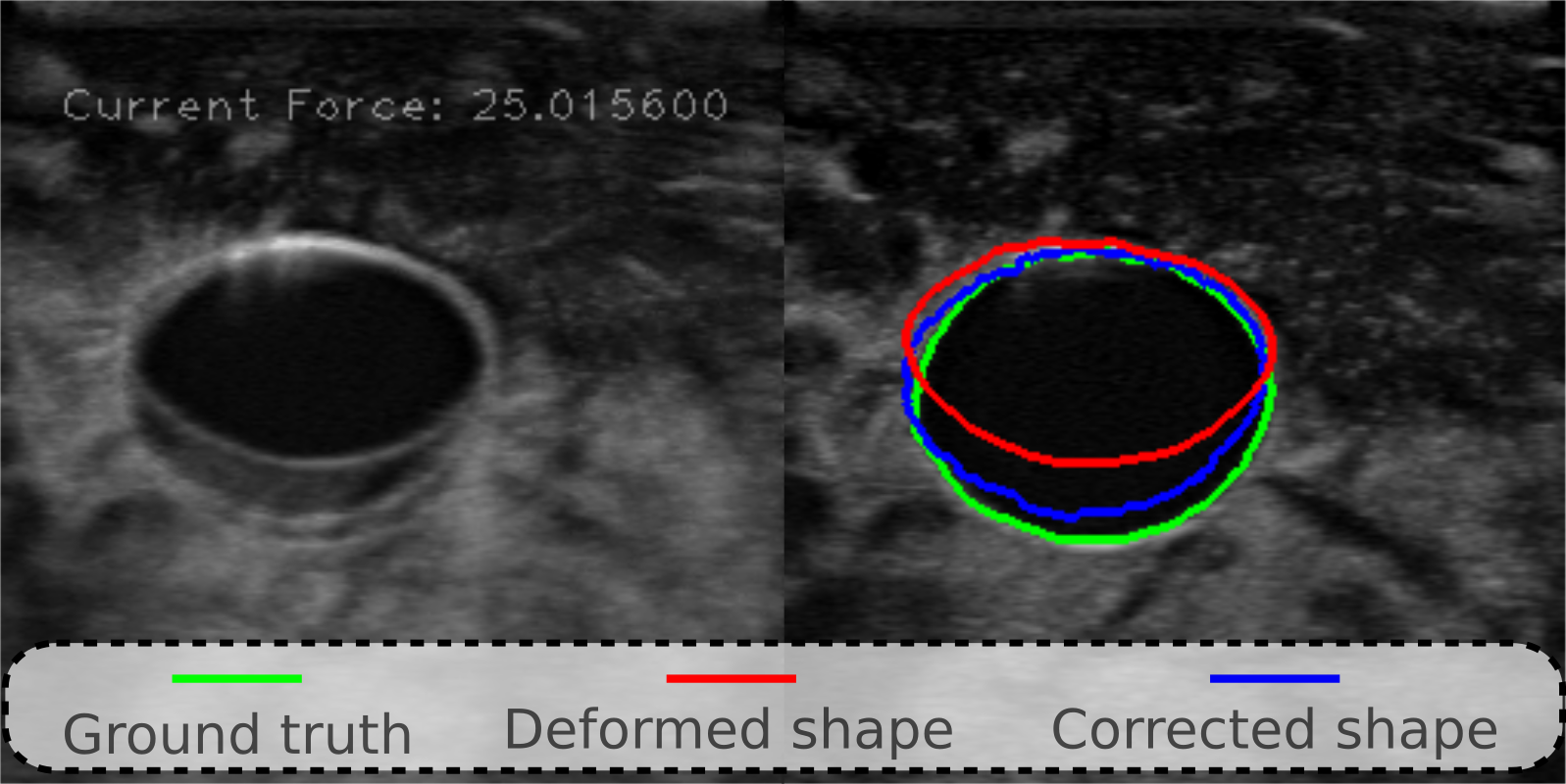}
\caption{Performance of the proposed deformation correction method for the 2D image obtained under $25~N$ on the stiff vascular phantom. The deformed image, corrected image and the ground truth acquired when the contact force is zero are overlapped in the left plot. The right image shows the extracted blood vessel boundaries.  
}
\label{Fig_2D_correction_A}
\end{figure}

The paired images, contact forces and probe displacements recorded from the third palpation at the same position are then used for validation. To demonstrate the correction performance, the results achieved when $F_c = 25~N$ is shown in Fig.~\ref{Fig_2D_correction_A}. The mimic artery geometry has been significantly compressed in the left view (Fig.~\ref{Fig_2D_correction_A}). This affects measurement accuracy of the object's geometry. Nonetheless, based on the results shown in Fig.~\ref{Fig_2D_correction_A}, the deformed geometry (red line) can be effectively recovered to ground truth (green line). To further quantitatively analyze the result, the dice coefficient is computed and it has been improved from $0.69$ to $0.92$.

\begin{figure*}[htb!]
\centering
\includegraphics[width=0.8\textwidth]{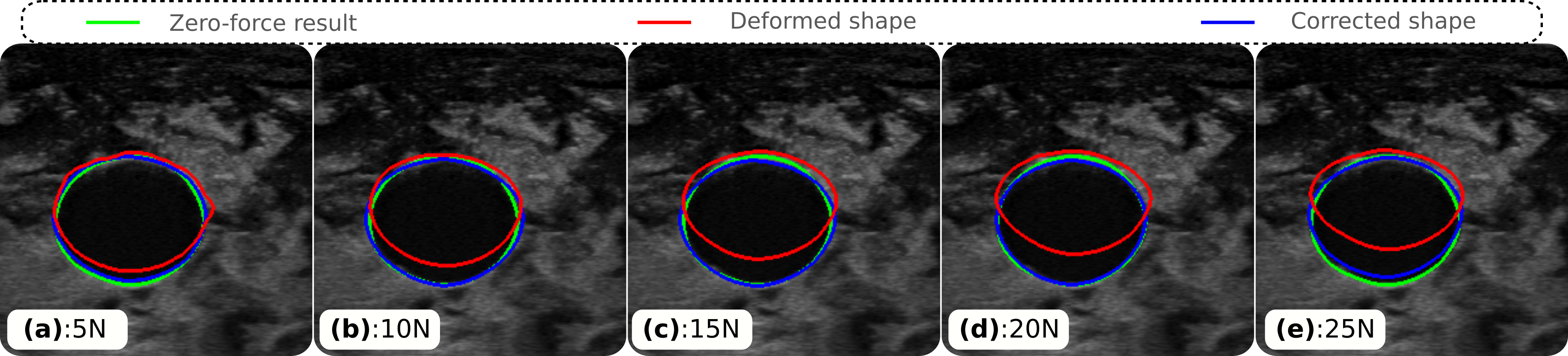}
\caption{The deformation correction performance on the same tissue at another position using the updated regression model based on local estimated stiffness. The computed dice coefficients for the deformed images and the corrected images are $[0.89, 0.87, 0.81, 0.78, 0.74]$ and $[0.97, 0.97, 0.97, 0.98, 0.94]$, respectively.
}
\label{Fig_2D_addptive}
\end{figure*}

\par
\subsubsection{Validation of Propagated Regression Model Computed based on Local Stiffness}
Considering the tissue properties are not the same at different positions, multiple palpation procedures are performed at two different positions (larger than $20~mm$ from the initial position) on the same gel phantom. Some results are shown in Fig.~\ref{Fig_2D_addptive}, which shows that the deformed tissue can be corrected using the updated regression model. The average dice coefficient for the corrected images is $0.96$. Besides, more results can be found in this video\footnote{https://www.youtube.com/watch?v=MlZtugQ2cvQ}.

\par
\subsubsection{Validation of Propagated Regression Model on Other Tissues}
\begin{figure}[ht!]
\centering
\includegraphics[width=0.35\textwidth]{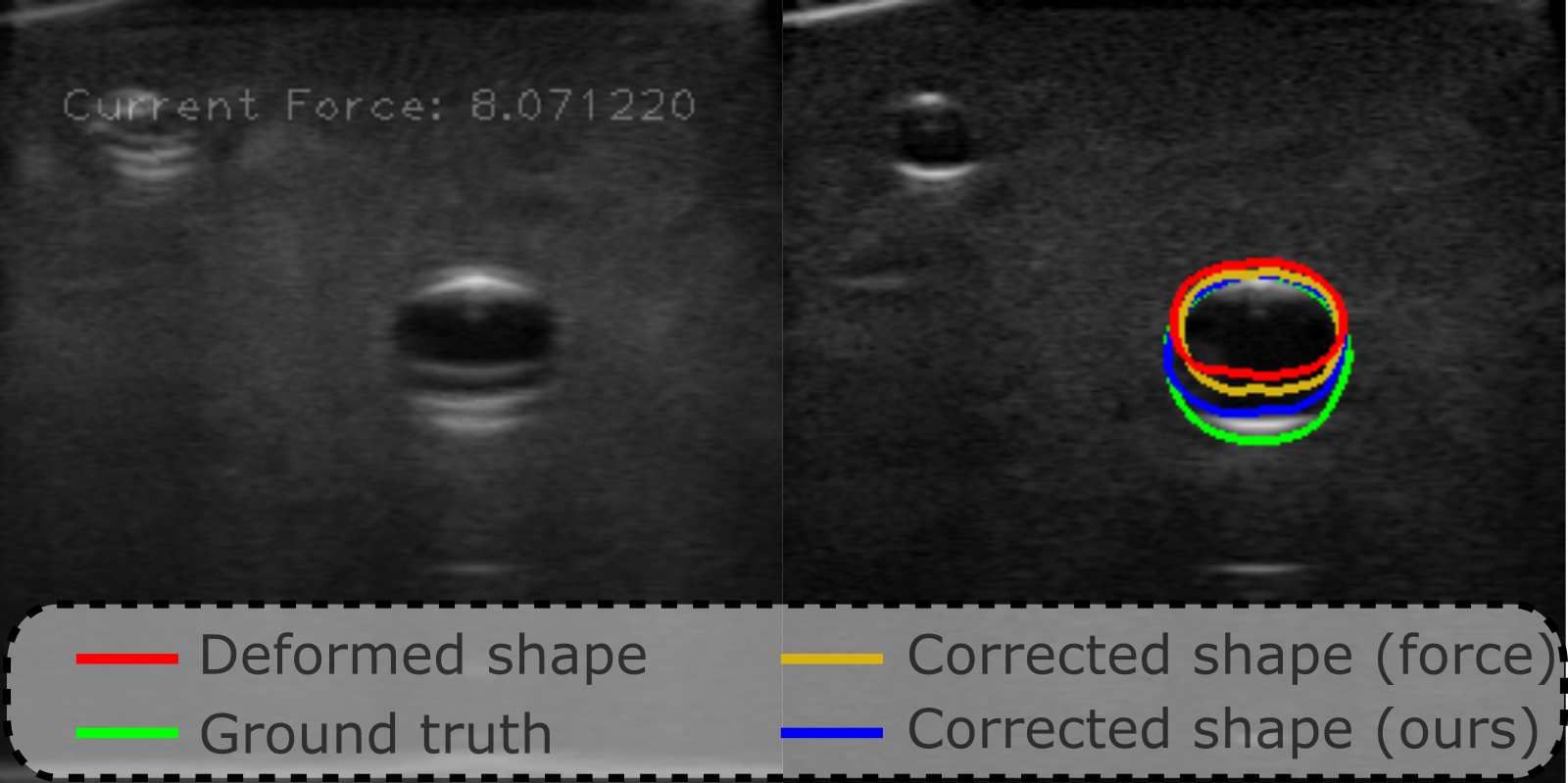}
\caption{Extrapolation performance on an ``unseen" soft phantom. Left: The overlapped result of the deformed image, ground truth and corrected images using the force-based approach~\cite{sun2010trajectory} and the proposed stiffness-based approach. Right: Vessel boundary detected on the four images. $F_c=8~N$. 
}
\label{Fig_2D_soft_addptive}
\end{figure}

To further validate the adaptive ability of the proposed method, a commercial vessel phantom is employed. The average stiffness of the soft phantom ($1489\pm617~N/m$) is only about half of the stiff one ($3237\pm56~N/m$). Besides, the stiffness of the soft phantom also varies much more than the stiff phantom when different contact forces are applied. This makes it challenging to adapt the optimized regression model based on the data acquired on the stiff phantom to an``unseen" soft phantom with such significantly different properties. The results are shown in Fig.~\ref{Fig_2D_soft_addptive}. The deformed vessel geometry (red ellipse) can be compensated as well using the proposed approach, and the corrected result (blue ellipse) is very close to the zero-compression data (green ellipse). The dice coefficients improved from $0.66$ to $0.89$ (i.e. by $35\%$). In addition, to demonstrate the advantages of the proposed stiffness-based approach over the existing force-based approach~\cite{sun2010trajectory}, the corrected result (yellow ellipse) obtained using the force-based approach is also shown in Fig.~\ref{Fig_2D_soft_addptive}. This result is obtained by substituting the force on ``unseen" tissues to an optimized quadratic polynomial regression model obtained for a sampled position on the stiff phantom. The corrected result of the propagated force-based model (dice: $0.74$) is significantly worse than the result of the proposed approach (dice: $0.89$).

\par
Finally, to systematically demonstrate the performance of the proposed method on different sampling positions both on stiff and soft tissues, the average dice coefficients for the deformed and corrected results have been summarized in TABLE~II. For the stiff phantom, four different sampling positions are employed to validate the performance of the adapted regression models at other positions. At each position, $F_c$ was gradually increased from zero to $25~N$. To demonstrate the correction performance under various force, the results obtained when $F_c = 5, 10, 15, 20, 25~N$ are summarized in TABLE~II.
Then, we further validate the method on the soft phantom. Since the phantom itself is softer, severe deformation happens when $F_c=8~N$ (Fig.~\ref{Fig_2D_soft_addptive}). Thus, the typical results acquired when $F_c = 2, 4, 6, 8~N$ over three sampling positions are summarized in TABLE~II. 

\par
\revision{In addition, the adapted results of force-based approach~\cite{sun2010trajectory} on the same points were computed as well (TABLE~II). Regarding the stiffness phantom, although the results of the proposed approach outperformed the results of the force-based approach in general, both of them achieved good performance. 
This is mainly because the stiff phantom is homogeneous, which resulting in similar stiffness for different positions on the same phantom. Since the stiffness of the soft phantom is really different from the stiff phantom, the adapted performance of the force-based approach becomes significantly worse than the results obtained using the proposed approach.
}

\begin{figure}[ht!]
\centering
\includegraphics[width=0.48\textwidth]{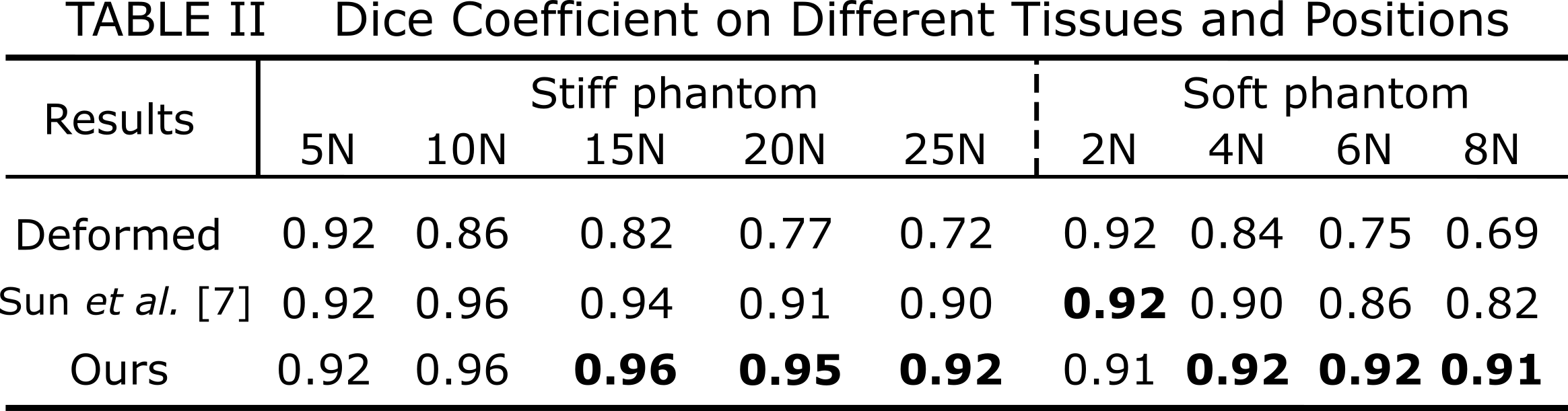}
\end{figure}

\subsection{3D Reconstruction Results}
\par
\begin{figure}[ht!]
\centering
\includegraphics[width=0.38\textwidth]{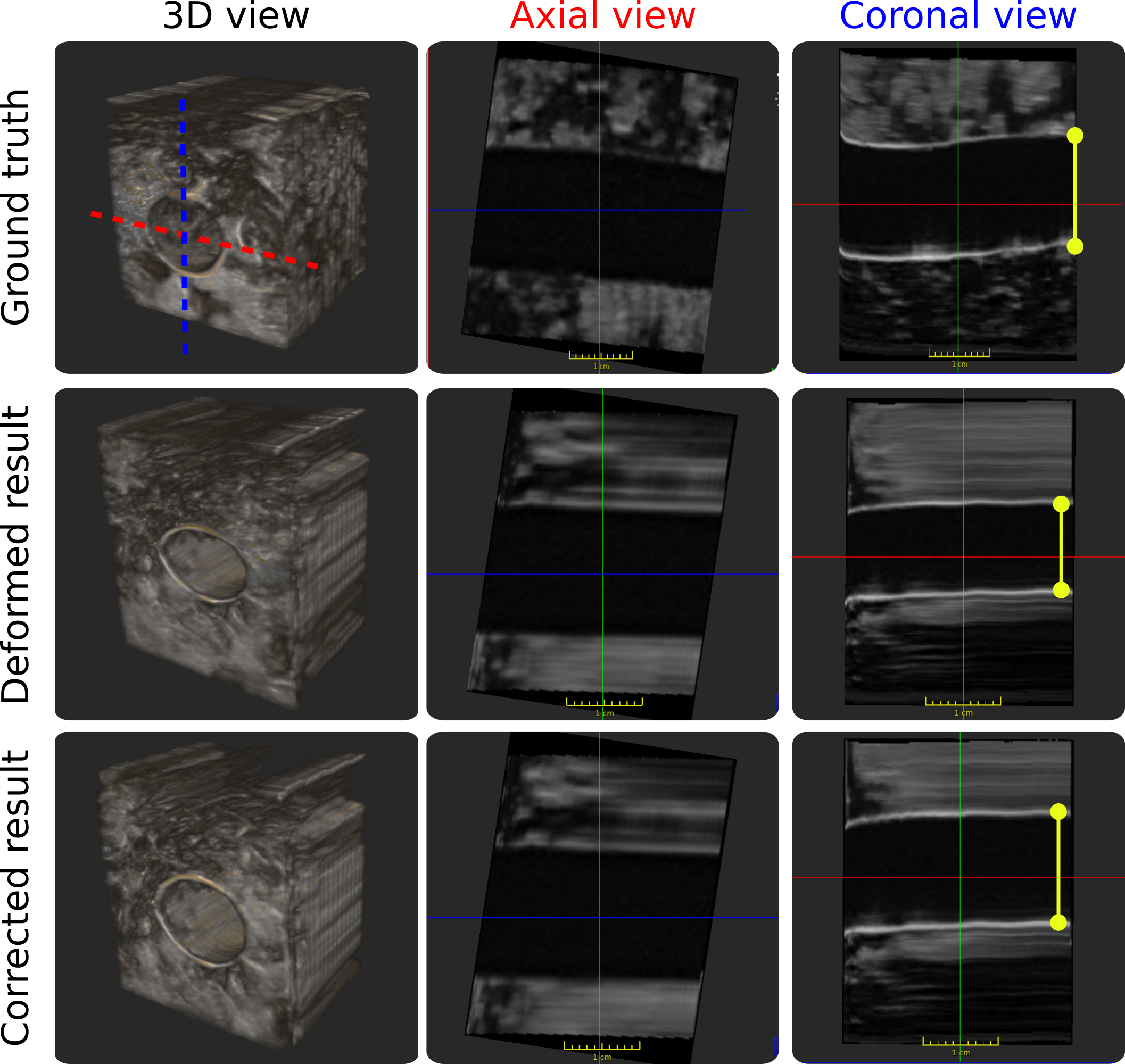}
\caption{3D compounding results. The ground truth is recorded when the contact force is zero (phantom is submerged in water). The deformed result is obtained when the force is $15~N$ on the stiff phantom. 
}
\label{Fig_3D_results}
\end{figure}

To obtain 3D volumes of objects, a sweep covering the area of interest is carried out. Considering the variation of tissue properties, $N_k$ sampling positions are selected. The dynamic stiffness at the sampling position is estimated based on the recorded $\lambda_z$. Here $N_k = 4$ is selected on the scan trajectory lengthened $40~mm$ for the stiff phantom while $N_k=3$ is used for the soft one, where the path is $60~mm$. To demonstrate the performance of the proposed stiffness-based regression model on the whole sweep, including many unsampled positions, the 3D volumes of the gel phantom (ground truth, deformed and corrected images) are shown in Fig.~\ref{Fig_3D_results}. The ground truth is obtained when the contact force is zero. To achieve this, the phantom is submerged in warm water to avoid air between the probe tip and object surface. 

\par
To better visualize and compare the resulting volumes, the views on the axial plane (dash red line) and coronal plane (dash blue line) are displayed in Fig.~\ref{Fig_3D_results}. 
The diameters of the three volumes measured in axial view are similar. However, in coronal view, the length of the orthogonal artery axis (yellow line) measured in the deformed results is less than $20\%$ of the length measured in the ground truth. After applying the proposed deformation correction method, the corrected results have successfully recovered the deformed coronal axis length from being almost identical to the length in the ground truth.

\par
\begin{figure*}[ht!]
\centering
\includegraphics[width=0.79\textwidth]{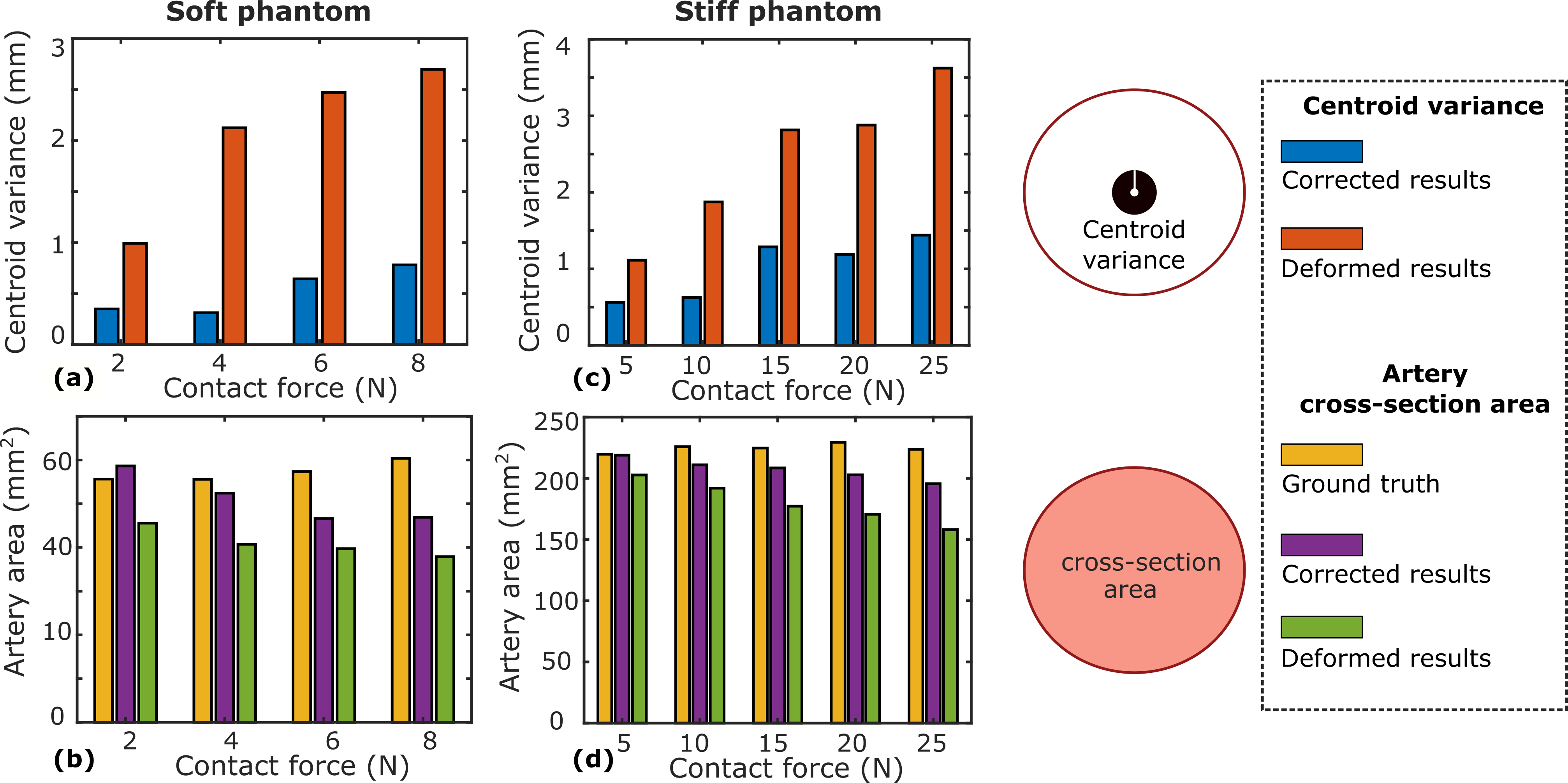}
\caption{(a) and (c) are the vessel centered variance between the ground truth and corrected and deformed results for the soft and the stiff phantoms, respectively. (b) and (d) are the vessel cross-section area for soft and stiff phantoms, respectively.  
}
\label{Fig_3D_results_bar}
\end{figure*}

To validate the accuracy and repeatability of the proposed robotic 3D US system, multiple robotic sweeps are carried out on the same trajectory. To generate the sweep with different levels of deformation, the US sweep is performed with various contact forces. Considering the phantoms' stiffness, five $F_c$ ($5, 10, 15, 20~\text{and}\ 25~N$) are used for the stiff gel phantom and four $F_c$ ($2, 4, 6, 8~N$) are used for the soft phantom. To statistically compare the correction performance in terms of force and stiffness, the vessel centroid variance and the cross-section area are introduced as the assessment metrics (see Fig.~\ref{Fig_3D_results_bar}).
The cross-section area is important because it is the main factor affecting the blood flow velocity and the flow volume pass in unit time. 
Besides, the centroid is computed for each B-mode image using OpenCV. Then, the centroid variance is computed as the Euclidean distances between the ground truth centroid and the centroids of the deformed and the corrected images. The variance only becomes small when both the long axis and short axis length of the vessel are close to the corresponding values in the ground truth. 

\par
Ten images are randomly selected from sweeps acquired under different $F_c$. The average results of the vessel centroid variance and the cross-section area are shown in Fig.~\ref{Fig_3D_results_bar}. It is noteworthy that the centroid variance is significantly reduced after correction in all investigated cases (different forces and phantoms), particularly for soft tissues. The variances for the corrected results are much smaller and more stable than the deformed cases. The maximum variances are $0.78~mm$ and $2.70~mm$ for the corrected and the deformed results on the soft phantom when $F_c = 8~N$. Regarding the stiff phantom, the maximum variances are $1.44~mm$ and $3.6~mm$ when $F_c = 25~N$. These values are larger than those peers obtained on the soft phantom. This is because the absolute geometry size of the artery in the stiff phantom is much larger than that in the soft phantom ($224~mm^{2}$ versus $57~mm^{2}$). 

\par
Regarding the vessel area, the corrected results are also closer to the ground truth than the deformed results. In the best-case scenario, $F_c$ is small for both phantoms ($2~N$ for soft phantom and $5~N$ for stiff phantom) because the pressure-induced deformation is also small when a small force is employed. However, in terms of the difference between the area of the ground truth and the corrected/deformed results, it is significantly reduced from $[17, 34, 47, 59, 66]~mm^{2}$ to $[1, 15, 16, 27, 28]~mm^{2}$ for the stiff phantom and from $[10, 15, 18, 22]~mm^{2}$ to $[-3, 3, 11, 13]~mm^{2}$ for the soft phantom. The improvement is over $55\%$ for all cases. Moreover, it can be seen from Fig.~\ref{Fig_3D_results_bar} that, the areas of the corrected results of the images acquired under different forces are very stable. The standard deviation between the areas of corrected images is only $8.8~mm^{2}$ and $5.7~mm^{2}$ for the stiff and the soft phantoms, respectively. 

\section{Discussion}
\revision{
The pressure-induced deformation is a common issue for US imaging of soft tissues. With the proposed approach, accurate and reproducible 3D images, independent of the experience of sonographers, can be achieved by correcting the US deformation. To qualitatively and quantitatively validate the proposed approach, blood vessel is investigated as the target anatomy. However, this study can benefit other applications requiring accurate geometrical measurements as well, such as examining and monitoring breast tumors~\cite{pheiffer2014model} and soft tissue sarcomas~\cite{virga2018use}.
Besides accurate 3D volumes, this study will also benefit the multimodal image fusion. The two typical clinical applications are image-guided intervention for soft tissues like breasts~\cite{ pheiffer2015toward, tagliabue2019position}, and imaging-guided orthopedic surgery~\cite{jiang2020automatic_tie}. For the former one, CT or MR is often used to provide high-resolution anatomies, while US images provide a live view during the intervention. Regarding the latter one, since patients may be moved after acquiring CT or MR images, it is necessary to do the registration between pre-operative images and patients by registering the live US images to the preoperative images. The corrected images could result in better results because it is easier to align the biological landmarks between the zero-compression B-mode images and preoperative images. 
}

\par
\revision{
The proposed deformation correction approach performs boldly in our setup of experiments. 
However, there are still} some limitations that need to be discussed.
First, the proposed method is only able to correct the deformations caused by the force applied on the object surface. \revision{This approach is not suitable for correcting deformations of deep organs, such as the liver. These deformations are caused by the coupling of applied external force and patient physiological movement like breathing.} 
In addition, to further improve the accuracy, the probe orientation should also be taken into consideration. This is not implemented here. Alternatively, one can apply methods \revision{that} enable the probe to be positioned perpendicular to \revision{the} surface \cite{jiang2020automatic}. Finally, in this work, we have so far considered the main in-plane deformation caused by the probe pressure on the surface. Such forces also result in residual out-of-plane deformations. However, since the main component of such deformation remains in-plane and the robotic ultrasound moves orthogonal to the B-mode acquiring all subsequent image planes, we consider this to be a valid approximation of the real tissue deformation.

\section{Conclusion}
In this work, we proposed a stiffness-based deformation correction method in order to achieve zero-compression 3D US images using robotic techniques. This method takes the nonlinear property of tissue stiffness as a key factor in correcting deformation. We obtained this patient-specified property by performing robotic palpation. Our approach enables the rapid adaptation of the optimized regression model to unseen positions by updating the local stiffness. Promising correction results have been achieved on both stiff and soft phantoms at arbitrary sampling positions. Additionally, the experimental results for 3D US acquired under different contact forces demonstrate that the proposed method is also able to recover zero-compression volumes from deformed images. This approach could even enable further techniques such as multi-modal image fusion. 

\section*{ACKNOWLEDGMENT}
\revision{
The authors would like to acknowledge the Editor-In-Chief, Associate Editor, and anonymous reviewers for their contributions to the improvement of this article. Besides, Z. Jiang wants to thank, in particular, the invaluable supports from his wife B. Zhang, the hero who just gave birth to their lovely daughter E. M. Jiang on 14.06.2021.}

\bibliographystyle{IEEEtran}
\bibliography{IEEEabrv,references}

\end{document}